                                 \documentstyle[12pt,aaspp4]{article}
\def\gtrsim{\mathrel{\hbox{\rlap{\hbox{\lower4pt\hbox{$\sim$}}}\hbox{$>$}}}}
\let\ga=\gtrsim
\def\lesssim{\mathrel{\hbox{\rlap{\hbox{\lower4pt\hbox{$\sim$}}}\hbox{$<$}}}}

\begin{document}

                                \title{
Prospects for Detection of Intracluster Gas Bulk Velocities Through the Sunyaev-Zel'dovich Effect
                                }

                                \author{
                                Renato A. Dupke \& Joel N. Bregman
                                }

                                 \affil{
Department of  Astronomy, University of Michigan, Ann Arbor, MI 48109-1090
                                }
                                \begin{abstract}

Intracluster gas velocity gradients have been recently detected by the authors 
in the Centaurus cluster (Abell 3526) using the 
Doppler shift of X-ray spectral lines with {\sl ASCA} Solid-state Imaging Spectrometers. 
The velocity gradient 
was found to be maximum along a line roughly perpendicular to the direction of the 
incoming subgroup Cen 45 and 
has a correspondent velocity difference of $\sim$ 3.4$\pm$1.1 $\times$10$^{3}$ km s$^{-1}$ within
a $\sim$10$^{\prime}$-diameter region centered on the cD galaxy NGC 4696. 
Such bulk velocities should Comptonize the Cosmic Microwave Background Radiation
producing variations of intensity and temperature that can be detectable in the near 
future with bolometers such as BOLOCAM . 
In this paper we realistically estimate the expected CMBR Comptonization for the central 
region of Abell 3526, using {\sl ASCA} and {\sl ROSAT} data to constrain the S-Z 
parameter expectations.

                                \end{abstract}


                                \section{
Introduction
                                }
				
Clusters of galaxies are believed to form from the infall/merging of smaller scale 
systems. Memory of the formation process may be probed through the X-ray analysis of the 
intracluster medium (ICM), where most of the cluster's visible mass resides. 
This is due to the fact that
the merging of sub-units is predicted to create gas substructures such as 
temperature and density inhomogeneities, destruction of cooling flows \& metal abundance 
gradients and generation of gas bulk velocities. The comparison of observed structures 
to numerical hydrodynamical+N body simulations can then provide clues as to the cluster's 
evolutionary stage (e.g. Evrard 1990; Katz \& White 1993; 
Navarro, Frenk \& White 1995; Evrard, Metzler, \& Navarro 1996; 
Roettiger, Burns \& Loken 1993,1996; Schindler \& Muller 1993; 
Pearce, Thomas \& Couchman 1994;  Gomez et al. 2001 and references therein).

Until very recently the physical characteristics of the ICM had been derived
almost exclusively by the analysis of distributions of electron temperature ($T_{e}$) 
and surface brightness ($S_{X}$) (or gas density),
even though in the last few years various numerical simulations of off-center 
cluster mergers have been published predicting the 
presence of long-lasting residual intracluster gas velocities approaching a few 
thousand km~s$^{-1}$ (e.g. Roettiger, Loken \& Burns 1997, Ricker 
1998; Roettiger, Stone, \& Mushotzky 1998; Takizawa \& Mineshige 1998; Burns et al. 1999; 
Takizawa 1999, 2000; Roettiger \& Flores 2000). Furthermore, since we are always looking 
at 2-D projections, velocity maps may be a crucial addition, necessary to break the degeneracies 
associated with analyses based solely on $T_{e}$ \& $S_{X}$ distributions. 

Recently, Dupke \& Bregman (2001a,b) detected ICM bulk motions in two galaxy clusters using 
{\sl ASCA} data: Perseus (Abell 426) and Centaurus (Abell 3526). The intracluster gas 
velocity distributions 
detected in these two clusters are roughly consistent with systematic gas bulk rotation 
with correspondent circular velocities $>$ 1000 km s$^{-1}$, implying that a significant 
fraction of the intracluster gas energy can be kinetic. 
Since X-ray measurements are haunted by the need of precise knowledge of instrumental gain
\footnote{conversion between pulse height and incoming photon energy}
these detections are significant only at the (2-4) $\sigma$ level. 
Realistically, the significance of these measurements are not expected to
improve substantially within the next several years even with {\sl Chandra} 
and {\sl XMM-Newton}, since the 
spectral resolution of the imaging spectrometers on-board these satellites, 
ACIS and EPIC\footnote{We are not considering the gratings
since they are less suited for this kind of spatially-resolved spectroscopy}, is only
slightly
better than that of the SISs on-board {\sl ASCA}. Furthermore, it will take a few years 
before detailed gain variations across those detectors are known well enough for reliable 
ICM velocity measurements. To compensate for the lack of knowledge of the gain 
variations in X-ray spectrometers it is necessary to take several consecutive off-center 
long-exposure observations of a cluster, so that sky regions with discrepant radial 
velocities can be observed at the same CCD position. However, this kind of measurement 
is very time consuming if one does not have prior knowledge of the best orientations 
for velocity measurements. 

Alternatively, ICM velocity measurements can be, in principle, 
corroborated/detected by the use of the kinetic S-Z effect 
(Sunyaev \& Zel'dovich 1970, 1972, 1980). 
Intracluster gas bulk velocities as
high as those detected in A3526 should 
generate significantly different levels of Comptonization of the cosmic microwave 
background radiation (CMBR)
towards different direction of the cluster (red-shifted and blue-shifted sides).
This effect could be 
detected with current (or in development) instruments, such as the 
{\sl BOLOCAM}\footnote{www.astro.caltech.edu/\~lgg/bolocam/bolocam.html}, or,
with smaller spatial resolution, 
{\sl ACBAR}\footnote{astrophysics.phys.cmu.edu/research/viper/} and the High Frequency Instrument 
on-board {\sl PLANCK}\footnote{astro.estec.esa.nl/Planck/}.

In this paper we realistically predict differential variation of the CMBR temperature
as it passes through the intracluster gas of the central regions
of the Centaurus cluster, $(\frac{\Delta T}{T})_{CMBR}$  
$\sim 10^{-4}$. We propose that this effect can be observed soon in cool 
clusters such as Centaurus at frequencies
that minimize the thermal component (214 GHz). We 
suggest that differential S-Z measurements of clusters, coupled with X-ray spatially 
resolved spectroscopy can provide crucial 
information not just to cosmology but also to the study of intracluster 
gas dynamics.

                                \section{
The Mechanism
                                }
The CMBR intensity and temperature
variations with respect to the background towards the direction of a rotating 
cluster has two components; a thermal and a kinetic component. The latter can be 
associated with some global peculiar velocity and also with internal
bulk motions, which is the one that we are interested in this work. If 
the resulting radial velocity of the kinetic S-Z component at some projected radius ``b'' is 
denoted by V$_{r}$ the thermal and kinetic variation of intensity ($I$) and temperature 
($T$) of the CMB in 
the non-relativistic approximation can be can be given by (for reviews see, e.g., Rephaeli 
1995; Birkinshaw 1999):

\begin{eqnarray}
\Delta I_{\nu} \approx & 2 \frac{(kT)^{3}}{(hc)^{2}} \frac{x^{4} e^{x}}{(e^{x} - 1)^{2}} (K(T_{e},\nu)
- \beta_{r}(b)) \tau(b)\;
\label{eq:adm}
\end{eqnarray}

\begin {equation}
(\frac{\Delta I}{I})_{\nu} \approx \frac{x e^{x}}{e^{x}-1} (K(T_{e},\nu) - \beta_{r}(b)) \tau(b)\;
\eqnum{2}
\end {equation} 

\begin {equation}
(\frac{\Delta T}{T})_{\nu} \approx (K(T_{e},\nu) - \beta_{r}(b)) \tau(b)\;
\eqnum{3}
\end {equation} 

where 

\begin {equation}
K(T_{e},\nu) = \frac{kT_{e}}{m_{e}c^{2}} (x \frac{e^{x} + 1}{e^{x} - 1} -4)\;
\eqnum{4}
\end {equation} 

\begin {equation}
\beta_{r}(b) = \frac{V_{r}(b)}{c}~and~x=\frac{h\nu}{kT}
\eqnum{5}
\end {equation} 

where $T_{e}$ \& $T$ are respectively the ICM and CMB temperatures, and the other 
parameters have
their usual meanings. If the gas 
number density $n(r)$ follows a king-like profile 
$n(r)= n_{0}(1 + (\frac{r}{r_{c}})^{2})^{- \frac{3}{2} \beta}$,
where ${r_{c}}$ and $n_{0}$ are respectively the core radius and the central density, 
the optical depth is given as a function of the projected radius ``b'' by

\begin {equation}
\tau(b) = \sigma_{T} n_{0} r_{c} B(\frac{1}{2} , \frac{3}{2} \beta - \frac{1}{2})(1 +
(\frac{b}{r_{c}})^{2})^{- \frac{3}{2} \beta + \frac{1}{2}} 
\eqnum{6}
\end {equation} 

where $B(p,q)= \int_{0}^{\infty} x^{p-1} (1+x)^{p+q} dx$ is the Beta function of p,q.

\section{The Case for Abell 3526}

Centaurus (Abell 3526) is a BM type I, nearby (z$\sim$ 0.0104), X-ray bright 
cooling flow (accretion rate $<$ 30-50 M$_{\odot} $yr$^{-1}$), cold 
(kT$_{e}$$ \sim$3-3.5 keV)) cluster. 
Its X-ray emission is relatively smooth except for the central 1$^{\prime}$ where
{\sl Chandra} found X-ray substructures  (Sanders \& Fabian 2001), and is peaked on 
the cD galaxy NGC 4696. The existence of galaxy velocity bi-modality in Centaurus 
had been shown by Lucey, Currie \& Dickens (1986a,b) and was more recently confirmed by Stein et al. (1997). 
Two galaxy groups are clearly separated: the main group (Cen 30), which is centered on the cD galaxy (NGC 4696) 
shows an average radial velocity of 3397$\pm$139 km~s$^{-1}$ and a velocity dispersion of 933$\pm$118 km~s$^{-1}$. 
The second group (Cen 45) is associated with the galaxy NGC 4709 at $\ga$15$^\prime$ from NGC 4696. 
It has an average radial velocity of 4746$\pm$43 km~s$^{-1}$ and a velocity dispersion of 131$\pm4$3 km~s$^{-1}$.
One arcmin at Centaurus distance corresponds to $\sim$ 19 h$_{50}^{-1}$ kpc. 
The general interpretation for this bi-modality is that Cen 45 is being accreted by the Centaurus cluster. This 
explanation has been further strengthened by: 1)the discovery of higher gas temperatures ($\ga$5 keV) 
associated with Cen 45 by 
Churazov et al. (1999) and 2) a marginally significant intracluster gas velocity difference of $\sim$ 1800 
km s$^{-1}$ between Cen 45 and Cen 30 detected by Dupke \& Bregman (2001b), hereafter DB01. 

More interestingly, DB01 found a small-scale (within the central 10$^{\prime}$) significant 
($>$99.8\% confidence) intracluster gas velocity gradient more or less symmetric with respect to the 
cluster's center,
consistent with bulk gas rotation with a correspondent circular velocity of 1.59$\pm$0.32 $\times$ 10$^{3}$ 
km~s$^{-1}$. The nature of this velocity gradient is likely to be related to some previous merger on the main body
of the Centaurus cluster (Cen 30). The maximum velocity difference is found roughly along an axis at a position 
angle of $\sim$ 40$^{\circ}$ (Figure 1). Therefore, velocity measurements by means of the kinetic S-Z effect 
on CMBR temperature and intensity variations could be optimized along that axis (e.g. 
by choosing symmetric subregions for source/background along that position angle, such as the extremes of 
the box in Figure 1).
 
In Figure 2 we show the absolute and relative variations of intensity and CMBR temperature towards
the directions of maximum (dotted) and minimum (dashed) velocities {\it as observed} by 
{\sl ASCA}, i.e., 5$^{\prime}$ away from the cluster's core (regions P3 \& P7 in Figure 1), 
as a function of wavelength. Both kinetic and thermal effects are added. The gas density
profile used was obtained taking into account that we are looking
at spatial regions with the same scale as (or smaller than) the cluster's core radius. Therefore, it is 
more appropriate to choose surface brightness profile fittings that realistically take into account 
the central cluster region, where the surface brightness is enhanced and single $\beta$ models
diverge significantly from the observed profiles. We believe that the double $\beta$ fitting profiles 
used in the extensive {\sl ROSAT} analysis 
of Mohr, Mathiesen \& Evrard (1999) (hereafter MME99) are the most reliable for that purpose and we use them 
in our calculations\footnote{The choice of MME99 is also a conservative one, since other surface-brightness-derived
density profiles available in the literature, e.g. Jones \& Forman 1999, typically produce higher values for the optical depth 
within the spatial regions considered, thus enhancing the magnitude of $\frac{\Delta T}{T}$.} 

MME99 surface brightness fitting function 
$S_{X}(b)=\sum_{i=1}^{2} S_{X0_{i}}(1 + (\frac{b}{r_{c_{i}}})^{2})^{- 3 \beta + \frac{1}{2} }$,
assumed a constant $\beta$ but different core radii and normalizations for the 
dominant components near the center and away from the central regions. To deproject
that surface brightness fitting and obtain a ``physical'' density profile MME99 expressed the 
total gas density as being proportional to the primary density component 
multiplied by a ``fudge'' function ($f(r)$) that compensated for the distortions caused by the 
central component. MME99 constrained $f(r)$ using the observed surface brightness
for each radial bin. 
Following their methodology we use for our $\tau(b)$ the following integral
\begin {equation}
\tau(b) =  2 \sigma_{T} \frac{n_{e0}}{f(0)} \int_{b}^{\infty} \nonumber\\
(1 + (\frac{r}{r_{c1}})^{2})^{-\frac{3}{2} \beta} \frac{f(r) r dr}{\sqrt{r^{2}-b^{2}}}
\eqnum{7}
\end {equation} 

instead of Equation 6. In Equation 7 $n_{e0}$ and $f(0)$ are the number density
and the ``fudge'' function ($f(r)$) at the cluster's center (MME99). We use for 
the functional form of $f(r)$ an inverse 
polynomial given as $f(r>1^{\prime}) \cong const0 + \frac{const1}{b} + \frac{const2}{b^{2}}$, where 
the best 
fit values within b$<$190 h$_{50}^{-1}$ kpc are $const0=0.817$, $const1=1.05 \times 10^{23} cm$ \& 
$const2=2.7 \times 10^{46} cm^{2}$. 

The surface brightness parameters used to derive the optical depth (Equation 7) are 
T$_{e}$(r=5$^{\prime}$) = 3.4 keV, V$_{circ}$(r=5$^{\prime}$)=1600 km s$^{-1}$ (DB01), $\beta$=0.57, 
$n_{e0}$=0.068 cm$^{-3}$, r$_{c1}$=7.3$^{\prime}$ (MME99).
In all three plots the solid thick line shows the difference between the 
redshifted and blueshifted curves.
It can be seen that the CMBR intensity variation difference (Figure 2, Top) is maximized at 214 GHz 
($\lambda$=0.14 cm). This is the optimal 
frequency to observe the relative kinetic S-Z effect and consequently to determine a velocity map.
At that frequency
the difference between the relative variations of intensity and temperature are
$\Delta(\frac{\Delta I}{I})\sim$ 5 $\times$ 10$^{-4}$ and 
$\Delta(\frac{\Delta T}{T})\sim  10^{-4}$ (Figure 2 middle and bottom plots). 

The magnitude of the parameters derived above are, naturally, a function of the
projected radial distance from the cluster's center. In order to determine the 
radial profile of the expected $\frac{\Delta T}{T}$ at 214 GHz we make two assumptions; 
Firstly,
we assume that in the region from 0$^{\prime}$ to 5$^{\prime}$ the cooling flow
can be approximated by a linear function varying from 3.4 keV to 1 keV at the 
center. The flattening of the temperature profile at the cluster's center has been 
observed for some clusters with {\sl Chandra} and {\sl XMM}, and may be a common feature 
of cooling flows (Fabian 2001). In the case of Centaurus there is evidence that the minimum 
temperature derived from spectral fittings using cooling flow models is $>$ 0.4-1.0 keV 
(Sanders and Fabian 2001). The central 
arcmin in Centaurus has an impressive amount of substructures (plumes) and this makes 
our estimation of the central densities more uncertain. Therefore, we exclude that region
from our analysis (Figure 3). Secondly, we assume the gas motion can be approximated
by a solid body rotation. This functional form for the gas velocity cannot be supported 
for larger radii since the gas becomes gravitationally unbound but it is a good
 observed first order approximation for the inner regions (Dupke \& Bregman 2002).
 Actually, this assumption is supposed to break at $\sim$ 5-6$^{\prime}$, when the 
 combined bulk kinetic and 
 thermal energies become greater than the gravitational and the gas becomes unbound 
 (e.g. see Allen \& Fabian 1994). Assuming that the velocity 
 ``freezes'' at 1600 km~s$^{-1}$, we also show in Figure 3 the predicted behavior of 
 $\frac{\Delta T}{T}$ at regions radially 
 greater than that encompassed by ASCA velocity analysis, for illustration purposes.
 
 The kinetic Comptonization of the spectrum that reaches us from a rotating cluster is 
roughly antisymmetric with respect to the center of the cluster from $\sim$ 160-270 GHz. 
This suggests that larger 
variations of intensity and temperature 
of the CMBR can be observationally enhanced by comparing the measurements (in the same channel) 
``within'' the 
cluster, rather than to a background region away from the cluster. The results are maximized at 214 GHz and we 
show the distribution of CMB temperature variations ($\frac{\Delta T}{T}(b)$) at that frequency in Figure 3.
The horizontal errors in Figure 3 show the FWHM of BOLOCAM. The solid lines represent the estimated
1-$\sigma$ errors of measurement associated to the X-ray-derived parameters, and are dominated
by X-ray velocity measurement errors. S-Z measurements of radial velocities can be performed within these
1-$\sigma$ limits with a relatively short exposure time (see below).

Variations of intensity or temperature of that magnitude can be relatively easily observed
currently, or in the very near future, with the new generation of bolometers. 
A total variation of CMB temperature 
of $\sim$ 0.3 mK between the ``knees''in Figure 3 is equivalent to
a flux density of $\Delta I_{lim} \sim$1.2 mJy~(FWHM)$^{-2}$, 
This could be done with high significance in a relatively short 
amount of time with, for example, {\sl BOLOCAM}, given that for that instrument the FWHM
is 43$^{\prime\prime}$ and the {\it expected}\footnote{Samantha Edgington and the 
Bolocam Team, Personal Comunication} noise equivalent flux density (NEFD) is
35~mJy~Hz$^{-\frac{1}{2}}$ at the 1.4mm band\footnote{The nominal observation 
time necessary to detect a limiting 
flux density $I_{lim}$ is $t_{obs}\sim (\frac{NEFD}{\Delta I_{lim}})^{2}$ sec plus overhead}(Glenn et al. 1998).

                                        \section{
Summary
}

Direct indications of the presence of large bulk motions in clusters of galaxies 
have been recently obtained through X-ray spatially resolved spectroscopy. However,
these observations are limited by detector's gain fluctuations and this will continue to
be the case for the next several years. An alternative way of measuring intracluster gas
bulk velocities is through the use of the kinetic S-Z effect. The use of bolometers has improved 
significantly the precision of radio observations of the S-Z effect, especially within
the optimal frequency range for such measurements, i.e., $\sim$ 214 GHz ($\lambda$ = 0.14 cm), 
so that, it will 
be possible in the near future to perform such velocity measurements 
in instruments such as {\sl BOLOCAM} (Bock et al 1998) with relatively short exposure
times. The combination of X-ray and radio measurements 
can provide us with independent/complementary information crucial to build 
ICM velocity maps and to determine the 
degree of evolution and the history of galaxy clusters.

So far, the best candidate for testing the proposed multi-wavelength measurements of 
velocity gradients is the Centaurus cluster. The symmetry of the 
gas velocity distribution and the low temperatures measured for this cluster should 
facilitate the separation of the rotational kinetic component from both thermal 
and peculiar velocity ones.
By combining the analyses of intracluster gas velocity \& temperature distributions
with other relevant X-ray cluster characteristics we predict 
that the variations of CMBR temperature between two regions symmetrically located 
5$^{\prime}$ from the clusters center along the line of maximal velocity gradient 
at $\lambda$ = 0.14 cm is $\sim$0.3 mK. 

The magnitude of the S-Z parameters observationally
derived in this work, if typical, could significantly enhance
possible halo-induced gas rotation contribution to the CMBR angular power spectrum 
(Cooray \& Chen 2002). However, the limiting nature of the search for 
velocity gradients using current X-ray spectrometers tends to be biased towards clusters
with high bulk velocities, so that the current data does not allow us yet to calculate 
precise velocity distributions and, consequently, to determine 
the deviations from the assumptions used in statistical analysis of the 
temperature anisotropies such as those performed by Cooray \& Chen (2002).
Furthermore, if both ``rotational'' and ``transient'' bulk velocities are as common 
in the central regions of clusters as suggested by a more extended ASCA analysis 
(Dupke \& Bregman 2002), measurements
of cluster peculiar velocities using the S-Z effect should be treated with extreme caution, given the 
potential confusion with an internal gas bulk motion component. 

Multiple off-center long exposure velocity measurements of intracluster gas 
with the {\sl Chandra} 
and {\sl XMM-Newton} satellites coupled with radio observations of the
differential kinetic S-Z effect 
will be crucial to determine precise intracluster velocities in the central
and intermediate regions of
galaxy clusters, improving significantly the determination of their 
evolutionary stage, and constraining biases related to the presence of 
coherent velocity fields in blind S-Z surveys.

\acknowledgments We are very grateful to Frits Paerels and Gary Bernstein for the 
helpful discussions and constructive suggestions.
We would like to thank Joe Mohr for kindly providing 
the tabulated distribution of the ''fudge'' function that was used in this paper.  
We also thank Samantha Edgington and the Caltech BOLOCAM team for helpful suggestions. 
We thank Hugh Aller \& Tim Paglione for helpful discussions. We acknowledge support 
from NASA Grant NAG 5-3247. This research made use of the HEASARC 
{\sl ASCA} database and NED. 

                                 
\clearpage
			      
                                \begin{figure}
                                \title{
Figure Captions
                                }
\caption{
Radial Velocity Distribution in Centaurus. (TOP) {\sl ASCA} GIS X-ray surface brightness
overlaid by the regions analyzed by Dupke \& Bregman (2001b). The values in parenthesis 
indicate the 1$\sigma$ confidence limits for one interesting paramenter (velocities). 
The box shows the region of maximum velocity gradient according to a solid body rotation
fit. "N" denotes the Northern direction. (Bottom) with an associated circular velocity of 1600 km s$^{-1}$. 
The vertical lines indicate the direction of the incoming subgroup Cen45. The two horizontal
lines show the 1$\sigma$ confidence limits for the velocity measured in the central 
region P0.
                                }
\caption{
Intensity and temperature variations as a function of wavelength for two
regions symmetrically located 5$^{\prime}$ from Centaurus' center along the direction 
perpendicular to the apparent rotation axis. The long-dashed \& dotted lines in all plots 
represent the curves for negative and positive values of radial velocity (V$_{r}$), respectively. 
The dashed-dotted line in the TOP plot shows the case for the thermal contribution alone 
(V$_{r}$=0). We also show the intensity variations for the kinetic component alone (T$_{e}$=0) 
in the TOP plot only, for illustration, and long-dashed \& dotted lines have the same meaning 
as described before. The solid line in all plots shows the magnitude of the 
difference between the negative and positive V$_{r}$ (differential variations). For intensity
variations (Top plot) this difference achieves a maximum at $\lambda$=0.14 cm ($\nu$=214 GHz).
                                }
\caption{
$\frac{\Delta T}{T}$ at $\lambda$ = 0.14 cm radial dependence along the direction of 
maximum velocity gradient. The solid lines indicate the 1-$\sigma$ errors correspondent to 
the ICM velocities measured by {\sl ASCA}. The horizontal errors indicate the FWHM of BOLOCAM 
for illustration purposes. The velocity profile is assumed to have a solid body form up 
to $\sim$ 5$^{\prime}$ and then to be constant at 1600 km~s$^{-1}$. The dotted line shows 
the expected temperature variation for the thermal component alone.
                                }
\end{figure}

                               
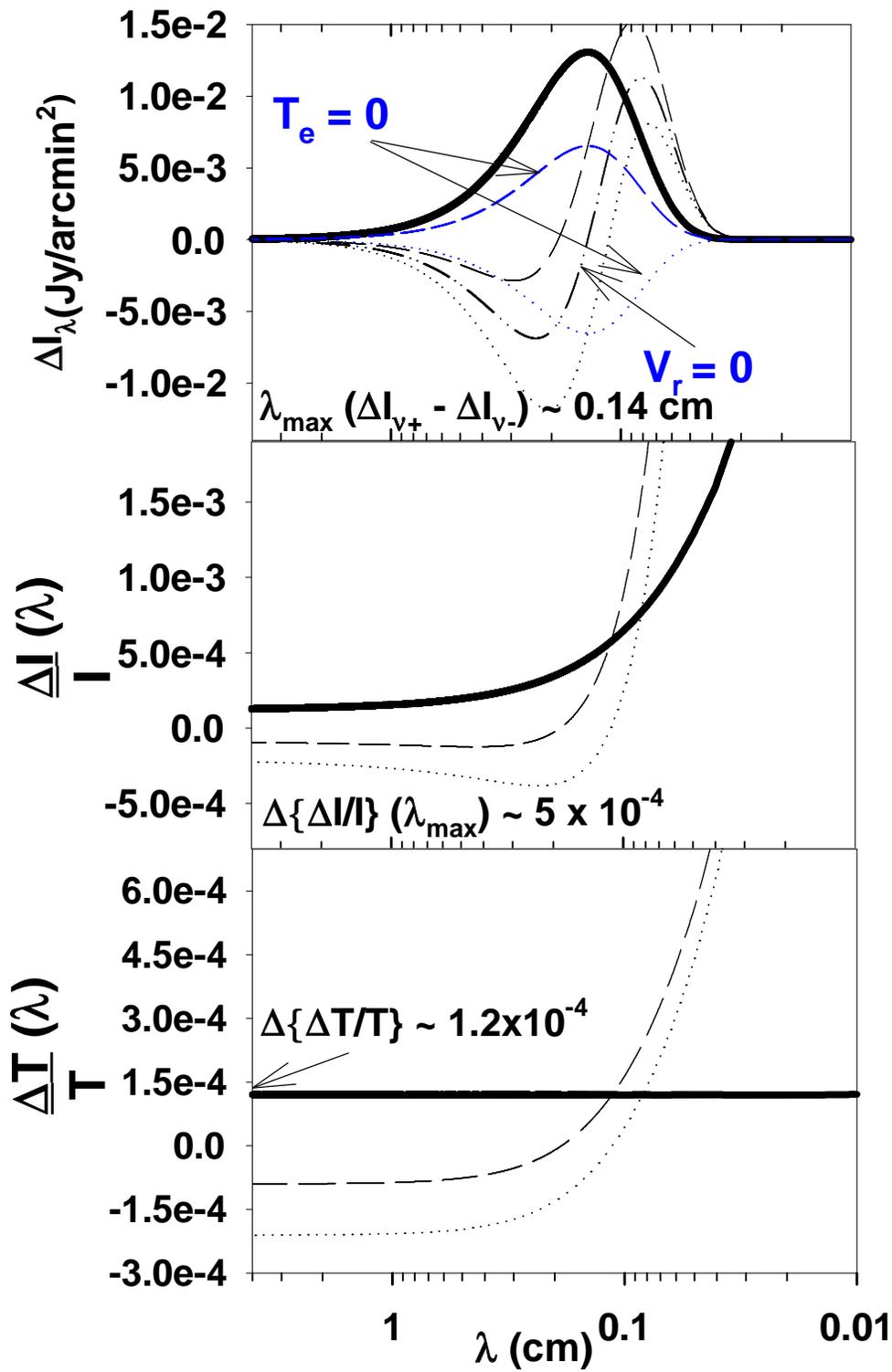
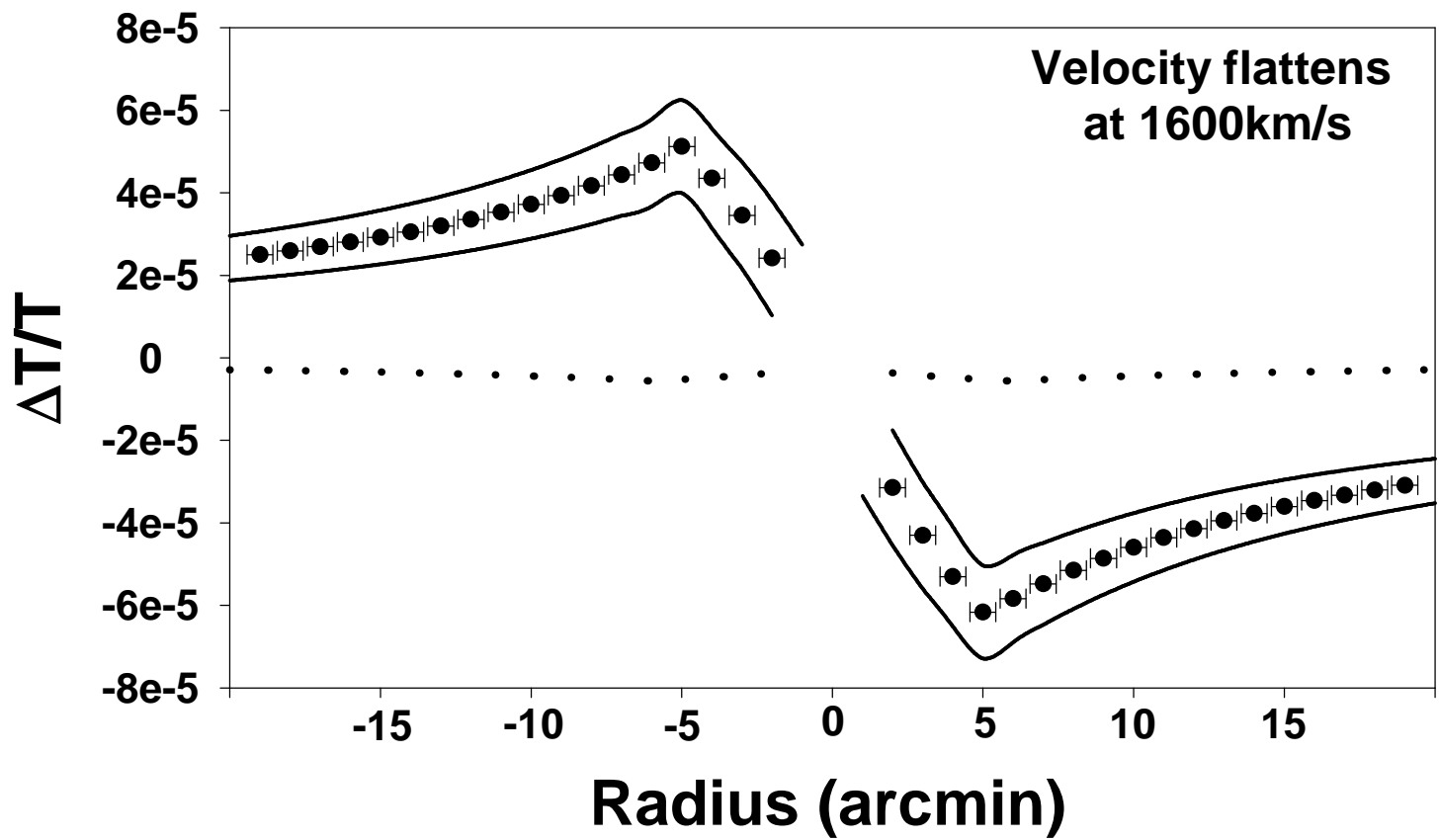
\end{document}